\documentclass[journal,a4paper]{IEEEtran}
\usepackage{url}
\usepackage[table,xcdraw]{xcolor} 
\usepackage{graphicx}
\usepackage[hidelinks]{hyperref} 
\hypersetup{
    colorlinks=true,
    urlcolor=black, 
    linkcolor=black, 
    citecolor=black 
}
\usepackage{multirow} 
\usepackage{float}
\usepackage{listings} 
\usepackage{style/json} 

\title{FLOWViZ: Framework for Phylogenetic Processing}

\begin{document}
\pagestyle{empty}
\author{Miguel Luís$^a$, Cátia Vaz$^a$\\
{$^a$DEETC, ISEL, Portugal}\\
\hspace{0.3cm} A43504@alunos.isel.pt \ cvaz@cc.isel.ipl.pt}

\maketitle

\begin{abstract}
  The increasing risk of epidemics and a fast-growing world population has
  contributed to a great investment in phylogenetic analysis, in order to
  track numerous diseases and conceive effective medication and treatments.

  Phylogenetic analysis requires large quantities of information to be
  analyzed and processed for knowledge extraction, using suitable techniques
  and, nowadays, specific software and algorithms, to deliver results as
  efficiently and fast as possible. These algorithms and techniques are
  already provided by several free and available frameworks and tools.
  Usually, the process of phylogenetic analysis consists of several processing
  steps, which define a pipeline. Some phylogenetic frameworks have available
  more than one processing step, such as inferring phylogenetic trees, data
  integration, and visualization, but due to the continuous growth in involved
  data amounts, each step may last several hours or days.

  Scientific workflow systems may use high performance computing facilities,
  if available, for processing large volumes of data, concurrently. But most
  of these scientific workflow systems cannot be easily installed and
  configured, are available as centralized services, and, usually, it is not
  easy to integrate tools and processing steps available in phylogenetic
  frameworks.

  This paper summarizes the thesis document of the FLOWViZ framework, which
  main goal is to provide a software integration framework between a
  phylogenetic framework and a scientific workflow system. This framework
  makes it possible to build a customized integration with much fewer lines of
  code, while providing existing phylogenetic frameworks with workflow
  building and execution, to manage the processing of great amounts of data.

  The project was supported by funds, under the context of a student grant of
  Funda\c{c}\~ao para a Ci\^encia e a Tecnologia (FCT) with reference
  UIDB/50021/2020, for a INESC-ID's project - NGPHYLO PTDC/CCI-BIO/29676/2017
  and a Polytechnic Institute of Lisbon project - IPL/2021/DIVA.
\end{abstract}

\vspace{.5cm}

\section{Introduction}
\label{section:introduction}

Biology has been strongly influenced by the digital era we live in. The
relationship between it and informatics had proven great humanitarian and
scientific advances, mainly in the areas of healthcare and virology, not only
contributing to create, but also enrich the science field of Bioinformatics.

Delivering fast and efficient results and track different viruses and diseases
is now a relevant world healthcare requirement, specially in the last two years
with the ongoing Covid-19 pandemic and due to the continuously growing world
population, which contributes to the appearance of new epidemics.

Consequently, scientists rely on phylogenetic and genome analysis for variant
tracking, in order to conceive effective medication.

Phylogenetic and genome analysis use large quantities of data, that needs to be
processed in order to be usable, which can sometimes reach terabytes (TB) in
size\cite{terabytes}. This added to the multiple tasks which compose various
analysis procedures, can make these very demanding, complex and time-consuming.
Specific software were made to ease the weight caused by these three issues,
where distributed and parallel computing are also used to decrease processing
times.

There are already multiple free and available solutions that let users make this
type of analysis, usually providing them with tools to build phylogenetic trees.
Most of them, however, only provide tools but do not provide ways to assemble
and create procedures with them. Because of this, scientists have to build these
procedures manually - they have to wait for results in order to proceed to the
next steps. Not only it is antiquated for today's standards, but also
time-consuming, inefficient and more prone to human error, as no automation is
involved.

These procedures are composed by groups of tasks or steps, which are usually
intrinsically dependent with each other - an output of a certain task may serve
as an input of a future task. This pattern repeats until the final task's
output, where the procedure ends. The procedure's structure and flow resembles a
pipeline or a flowchart, being this the reason why the scientific community
labeled these procedures as \textit{pipelines} or \textit{workflows}.

To automate these procedures, \textit{Scientific Workflow Systems} were
developed - specialized software, which provide a Domain-Specific Language
(DSL), that allow users to script their own workflows and manage complex
distributed computation and data in distributed resource environments.

Nowadays, there are many available workflow systems and each one provides a
different and, sometimes, unique way to build workflows. At first, this software
diversity can bring many options to build workflows, however, workflow
shareability worsens, as there are many DSL to configure the same workflow for
different workflow systems.

In order to provide workflow shareability among different workflow systems, the
\textit{Common Workflow Language} (CWL)\cite{cwl} standard was created. This new
standard contributed to an increasing interoperability between different
workflow systems, however, as it is still new, a small percentage of them
support this standard. It is expected that, given this standard increasing
popularity, more workflow systems will implement it.

However, scientific workflow systems are not easy to configure and most of them
are available through centralized services. These implementations also preclude
the user from including new phylogenetic tools and hamper additional and
customized configurations. It would also be easier if the user could update
their phylogenetic frameworks to a version that supported pipeline specification
and execution automatically, through workflow systems, by simply installing a
package and applying a small configuration to it.

The main goal of this paper is to propose an architecture which is being applied
in a developing framework - \textit{FLOWViZ}, that aims to offer workflow
specification and execution to existent phylogenetic frameworks, with minimal
configuration. By offering an user-friendly web interface, where users can
manage and build workflows, allowing them to add and use their own tools, along
with the bundled ones inside the framework itself. This is achieved by letting
users define \textit{contracts} - interfaces for their tools, which specify
rules and operation guidelines for a correct tool execution and configuration.
This way, the framework gets to know how to invoke a tool and how it should be
correctly configured by the user, allowing the framework to integrate new tools,
without changing its source code and without requiring the integrating tool to
adapt. The framework, which is composed by a client and a server, relies on a
workflow system, that schedule the workflows' execution and return the results
back to the framework, which will be latterly delivered to the user.

With this approach, the developing framework will provide: \textit{(i)
    Automation} - make the phylogenetic analysis an automatic process;
\textit{(ii) Flexibility} - allow the user to implement a wide range of
phylogenetic tools, which can be executed in many execution environments,
such as containers or cloud virtual machines; \textit{(iii) Scalability} -
support large-scale analysis, by relying on a workflow system that supports
distributed and parallel computation in large clusters; \textit{(iv) Result
    production} - provide complete and detailed results and logs, regarding the
workflow's execution; \textit{(v) Interoperability} - allow seamless
integration with other phylogenetic frameworks that want to provide workflow
building, through contracts and components' loosely-coupled relationships;
\textit{(vi) Reproducibility} - supply ready-to-use tools and workflows,
which allow users to reproduce the same procedures; allow them to share
their workflows through the CWL standard.

The rest of the document is divided by as follows. Section \ref{relWork} shows
the studied state-of-the-art phylogenetic frameworks and workflow systems.
Section \ref{overview} presents the architecture of the project, namely the main
components and interactions among them. Sections \ref{toolIntegrationSec},
\ref{workflowBuildingSec} and \ref{resultProductionSec} refer and expose details
of specific system's modules.

Finally, in section \ref{discussion} we discuss the proposed framework, expose
possible future work and summarize the main contributions.
\section{Related Work}
\label{relWork}

This section presents all the studied related work, namely, state-of-the-art
phylogenetic frameworks and workflow systems.

\subsection{Phylogenetic frameworks}

We studied two well-known phylogenetic frameworks: PHYLOViZ\cite{phyloviz} and
NGPhy\-logeny\cite{ngphylogeny}.

PHYLOViZ provides methods to build phylogenetic trees in real-time, through a
user-friendly interface. It also provides other functionalities, such as
distance matrix visualization and sequence visualization. However, phylogenetic
procedures need to be manually built, one step at a time, not providing
functionalities to build workflows.

On the other hand, NGPhylogeny is a phylogenetic web framework, which is
integrated with the Galaxy\cite{galaxy} workflow system. The framework provides
three ways to build workflows: (1) using pre-made workflows with default values;
(2) using pre-made workflows with users' values; (3) building customized
workflows, using the framework provided tools and users' inputs. This method is
also labeled as \textit{"à la carte"} workflows.

Although NGPhylogeny provides users with workflow building and scheduling, they
are limited to the tools that the framework has to offer. This is a problem that
our framework aims to solve, by allowing users to integrate their tools and use
them along with the ones that came bundled with the integrating tool. This can
be achieved by contract specification, where rules and guidelines of each
integrating tool are specified.

\subsection{Workflow systems}

In the context of this work, the \textit{workflow} is a generic term to
designate the automation of a process, in which data is processed by different
logical data processing activities according to a specified set of rules
\cite{workflowSurvey}. It can also be interpreted as a \textit{flow of work}
that have a beginning and an end. Moreover, in Bioinformatics, it is usual to
use the term workflow or pipeline for expressing the same.

As workflow scheduling and execution is the core requirement of this project,
workflow systems were researched, in order to find the one that would better fit
the framework's requirements. We previously selected three well known scientific
workflow systems, namely: \textit{Apache Airflow}\cite{airflow},
\textit{Next\-flow}\cite{nextflow} and \textit{Snakemake}\cite{snakemake}.

The selection was based on their popularity among the scientific community and
with workflow systems which have a programming base language.

We analyzed the relevant features of them, which are summarized in Table
\ref{compTable}.
\begin{table}
    \centering
    \caption{Workflow comparison table}
    \label{compTable}
    \footnotesize
    \begin{tabular}{|c|ccc|}
        \hline
        \multirow{2}{*}{\textbf{Characteristics}}                                                          & \multicolumn{3}{c|}{\textbf{Workflow System}}                                                               \\ \cline{2-4}
                                                                                                           & \multicolumn{1}{c|}{\textbf{Airflow}}         & \multicolumn{1}{c|}{\textbf{Nextflow}} & \textbf{Snakemake} \\ \hline
        \textbf{Base Language}                                                                             & \multicolumn{1}{c|}{Python}                   & \multicolumn{1}{c|}{Java / Groovy}     & Python             \\ \hline
        \textbf{\begin{tabular}[c]{@{}c@{}}Composition Style\\ (Values: Script, GUI)\end{tabular}}         & \multicolumn{1}{c|}{Script}                   & \multicolumn{1}{c|}{Script}            & Script             \\ \hline
        \textbf{\begin{tabular}[c]{@{}c@{}}Execution Style\\ (Values: CLI, GUI)\end{tabular}}              & \multicolumn{1}{c|}{GUI / CLI}                & \multicolumn{1}{c|}{CLI}               & CLI                \\ \hline
        \textbf{\begin{tabular}[c]{@{}c@{}}CWL Support\\ (Values: Yes, No)\end{tabular}}                   & \multicolumn{1}{c|}{Yes}                      & \multicolumn{1}{c|}{No}                & Yes                \\ \hline
        \textbf{\begin{tabular}[c]{@{}c@{}}Containerization\\ Support\\ (Values: Yes, No)\end{tabular}}    & \multicolumn{1}{c|}{Yes}                      & \multicolumn{1}{c|}{Yes}               & No                 \\ \hline
        \textbf{\begin{tabular}[c]{@{}c@{}}Execution Order\\ (Values: Explicit,\\ Implicit)\end{tabular}}  & \multicolumn{1}{c|}{Explicit}                 & \multicolumn{1}{c|}{Implicit}          & Implicit           \\ \hline
        \textbf{\begin{tabular}[c]{@{}c@{}}Dependency Order\\ (Values: Explicit,\\ Implicit)\end{tabular}} & \multicolumn{1}{c|}{Explicit}                 & \multicolumn{1}{c|}{Implicit}          & Implicit           \\ \hline
        \textbf{\begin{tabular}[c]{@{}c@{}}Workflow Sharing\\ (Values: Yes,\\ Partial, No)\end{tabular}}   & \multicolumn{1}{c|}{Yes}                      & \multicolumn{1}{c|}{Partial}           & Partial            \\ \hline
    \end{tabular}
\end{table}
One of the requirements is containerization support. As we can observe in Table
\ref{compTable}, Snakemake does not support this core feature and, therefore, we
discarded this workflow system. Another core feature is CWL support, which
Nextflow does not have and, thus, this workflow system was also discarded. The
workflow system must also provide workflow composition via CLI, which is
supported by all of them. Execution and dependency order are not core features,
however, explicit order is more manageable than implicit and thus preferred, as
it allow tasks' order and data dependencies to be defined explicitly, and not
only by the tasks' topology: in which order tasks were written inside the
script.

Given this, the workflow system which gathered most benevolent characteristics
and is the most suitable for this project is \textit{Apache Airflow}: it
supports container execution and has CWL support. It also allows to specify the
execution order of tasks explicitly and it also provides complete logs of the
executed tasks, via GUI or CLI.
\section{FLOWViZ Overview}
\label{overview}
FLOWViZ\footnotemark[1]  is a software integration framework with the goal of
providing a bridge between an existing phylogenetic framework and a workflow
system. By supplying a basic configuration with few lines of code, the user can
seamlessly integrate customized phylogenetic tools, which can be used to build
and execute workflows and deliver results that better fit users' needs.

As a case study, we chose to use the PHYLOViZ framework, as the phylogenetic
framework, and Apache Airflow\cite{airflow}, as the workflow system. We chose
PHYLOViZ since it is a well known phylogenetic framework in this field and
provides a great set of phylogenetic tools, however, it lacks workflow building
and execution, making this a relevant case study for this implementation.

FLOWViZ has four primary requirements: (i) allow the user to integrate new tools
with any phylogenetic framework and use them along with priorly added tools;
(ii) easy integration with the phylogenetic framework, by supplying visual
elements to easily access FLOWViZ features; (iii) manage and create workflows;
(iv) export created workflows through CWL-written scripts, in order to
facilitate workflow shareability;

\footnotetext[1]{The public code repository of FLOWViZ is available through this
    link:\\
    \href{https://github.com/mig07/FLOWViZ}{https://github.com/mig07/FLOWViZ}}

To achieve a seamless tool integration, the proposed framework allow the users
to establish contracts, where the tool's rules and guidelines are specified.
This also happens when bridging the phylogenetic framework with FLOWViZ, namely
the developer or the person responsible for the frameworks' integration must:
(1) - have the phylogenetic framework's tools exposed via API or CLI; (2) -
build the contracts to make these tools usable by the FLOWViZ framework.
Customized tools added by users, must be deployed in a remote computing instance
(cloud virtual machines or containers), with the necessary open ports and
security configurations, to make the tool accessible by FLOWViZ. After this, the
tool contract can be established and the tool can be successfully invoked by the
framework.

\begin{figure}
    \centering
    \includegraphics[width=\linewidth]{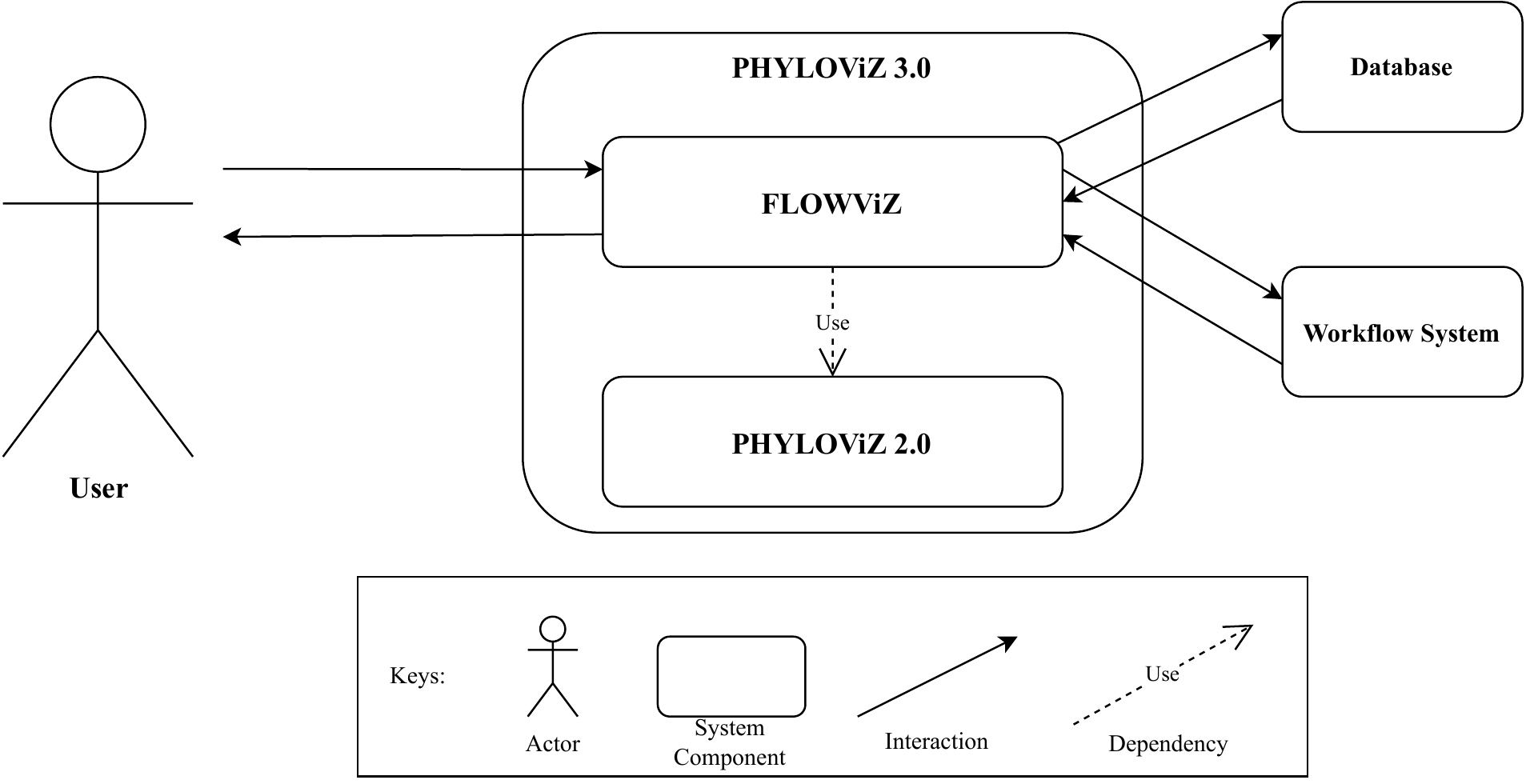}
    \caption{System's architecture}
    \label{arch}
\end{figure}

Self-hosting FLOWViZ does not require exposing tools using remote instances - in
\textit{localhost} environment, the user or developer only needs to specify the
tools' contracts. FLOWViZ will be a component of a system that integrates the
phylogenetic framework and the workflow system. Therefore, the system
architecture is composed by \textit{four} main components: the \textit{FLOW\-ViZ
    framework}, the \textit{phylogenetic framework} (in this case, PHYLOVIZ 2.0),
the \textit{workflow system} and the \textit{database}. Figure \ref{arch} show
the system architecture.

%
%
In this architecture, integrating FLOWViZ with PHYLOViZ 2.0, will generate a new
version of the PHYLOViZ project - version 3.0.

The system's architecture can also be divided by three functional modules: {\it
(i) integration module}; {\it (ii) workflow building module}; {\it (iii) result
production module}. The {\it integration module (i)} is responsible for
integrating new phylogenetic tools within the framework via contracts; the {\it
        workflow building module (ii)} refers to the part responsible to supply a GUI
that allows workflow building and scheduling; the {\it result production (iii)}
uses the workflow execution logs to build a final report related to the workflow
execution.

The system assets are the visual representation of the data, which is used
during the workflow execution: tools and necessary files.

\begin{figure}
    \centering
    \includegraphics[width=\linewidth]{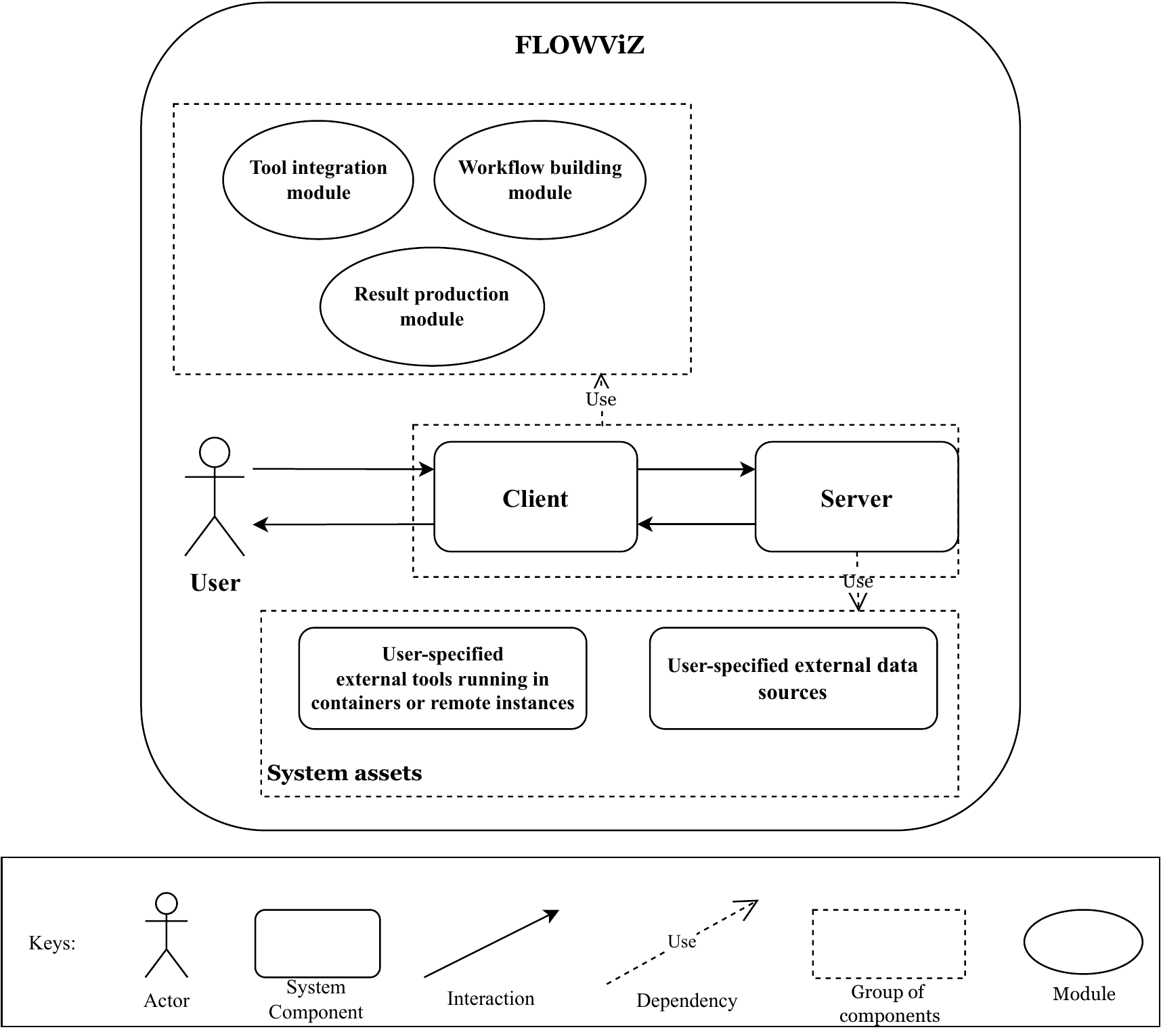}
    \caption{FLOWViZ architecture detail}
    \label{flowvizArch}
\end{figure}

The FLOWViZ component encapsulates two sub-components: the \textit{client} and
the \textit{server}. In the end, the FLOWViZ component serves as a middleware
among: the user, the workflow system and the phylogenetic framework. This can be
observed in Figure \ref{flowvizArch}.
%
%
As presented in Figure \ref{flowvizArch}, the FLOWViZ component is composed by:
a web client written in JavaScript with React, which supplies the user with a
graphical user interface allowing it to add its own tools, build workflows and
display results; an HTTP Express server written in JavaScript and Node.js, which
supplies all the necessary endpoints to the client; it also provides user
authentication by communicating with the phylogenetic framework; a database
where all user-defined contracts and other relevant metadata are stored; the
system assets, which are only the visual representation for the tools and
necessary files for workflow execution; the functional modules that this
component performs: \textit{tool integration (i)}, \textit{workflow building
    (ii)} and \textit{result production (iii)} modules.

The following sections detail each system's modules, approaching the respective
involved components and specific use cases.
\section{Tool integration module}
\label{toolIntegrationSec}
\begin{figure}
    \centering
    \includegraphics[width=\linewidth]{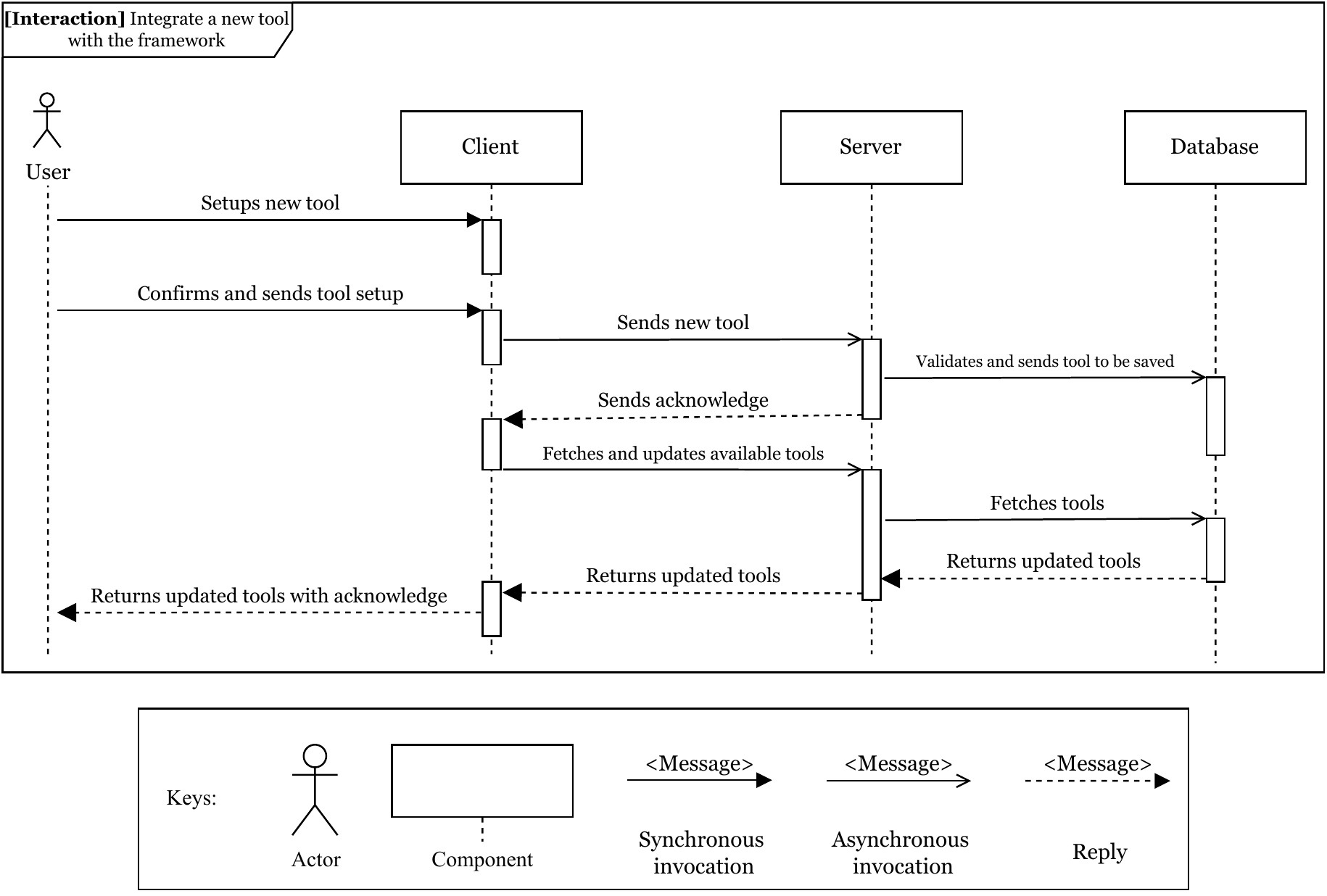}
    \caption{Interaction diagram: tool integration module's general use case}
    \label{integration}
\end{figure}

The \textit{integration module (i)} is responsible for enabling the user to integrate new
tools inside the framework, which is the contract establishment for each tool. Figure
\ref{integration} shows the interaction diagram of the integration module's
general use case.
In the depicted interaction diagram, Figure \ref{integration}, the
user configures the new tool using the provided web client GUI.
When the setup is concluded and sent to the server, this receives, validates it and, finally, saves the contract
into the database. Afterwards, the client self updates with the newest provided
tools and receives the acknowledgment signal
that the added tool was successfully integrated with the framework.

There are two ways to configure a tool: if it
has CLI support, the user can configure it as a library and configure each
command's invocation and settings; or if the tool exposes an API, the user can
set up each exposed endpoint, providing the allowed structure for the HTTP body and headers.

Listing \ref{toolContract} shows the generic tool contract. This listing has
both \texttt{api} and \texttt{library} fields for demonstration proposes but,
in a real scenario, only one of them should be defined, since the user previously
decides if it wants to configure the tool access as a library or as an API . This happens because the user
firstly decides if it wants to configure the tool access as a library or as an
API.
\begin{lstlisting}[language=json, label={toolContract}, caption={Generic tool contract (updated september, 2022)}]
{
  "general": {
    "name": "Name of the tool.",
    "description": "Tool description."
  },
  "access": {
    "_type": "type of tool access (library or API)"
    // External tool access specifications
  },
  "library": [
    // Groups of commands
  ],
  "api": [
    // Endpoints
  ]
}
\end{lstlisting}
\section{Workflow building module}
\label{workflowBuildingSec}
\begin{figure}
    \centering
    \includegraphics[width=\linewidth]{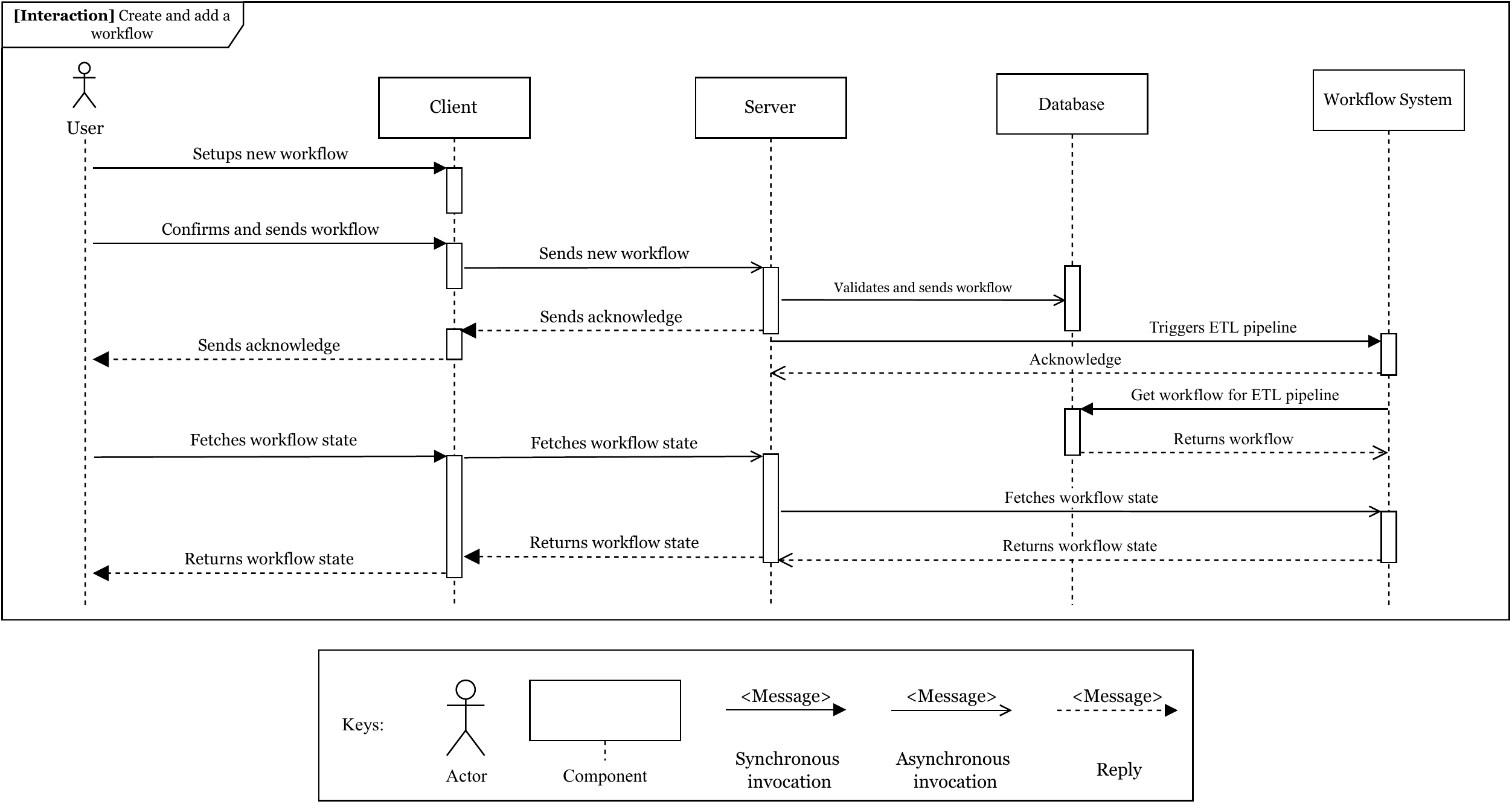}
    \caption{Interaction diagram: workflow building module's general use case}
    \label{workflowBuilding}
\end{figure}

The \textit{workflow building module (ii)} provides the user with a graphical
user interface to build their workflows, using the user provided tools and the
ones which came already bundled with the framework. The GUI consists of a
whiteboard with a tools' side list, which the user can use to \textit{drag and
    drop} tools into the whiteboard and connect them, creating a flowchart and
defining the workflow. Figure \ref{workflowBuilding} shows the interaction
diagram of the workflow building module's general use case.
As shown in Figure \ref{workflowBuilding}, after the tool's configuration,
the user can build a workflow, by using the provided GUI, the whiteboard,
where the workflow can be graphically drawn and each invoked tool setting can be individually configured.
When the workflow configuration is concluded, the user submits it and the client sends the workflow
to the server, which will go through a validation, that checks if every tool configuration is valid.
If the validation succeeded, the workflow will be saved into the database and a
HTTP request, containing the workflow's name and the correspondent user, will be
sent to the Airflow's REST API, which will fetch the submitted workflow and
translate its data into an Airflow DSL script, also known as an DAG (Directed
Acyclic Graph). The DAG will then be executed at the date and time,
pre-configured by the user, and the workflow's execution will produce results.
\section{Result production module}
\label{resultProductionSec}

\begin{figure}
    \centering
    \includegraphics[width=\linewidth]{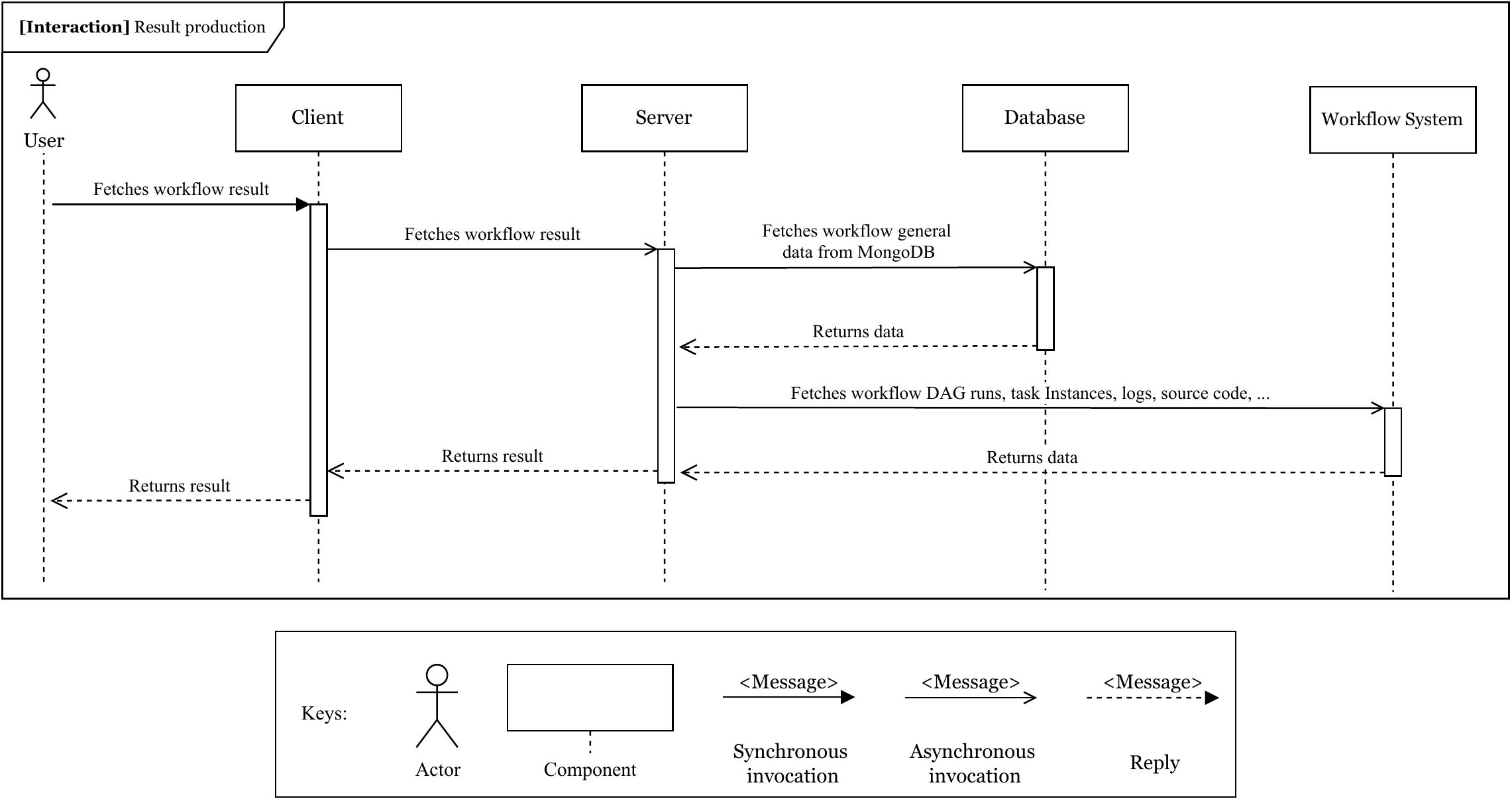}
    \caption{Interaction diagram: result production module's general use case}
    \label{resultProduction}
\end{figure}

The \textit{result production module (iii)} delivers the workflow log back to
the client, so the user can retrieve it. When the workflow execution finishes,
the HTTP server provides the client with endpoints, which fetch data from the
Airflow REST API, in order to retrieve results for a specific workflow. Figure
\ref{resultProduction} displays the general use case implementation of the
result production module. When the workflow's execution finishes, the user can
fetch its results via web client, which will send a request to the HTTP server.
This will fetch data from both the database and the Airflow's REST API and will
send it back to the client. These results include logs regarding each workflow
execution, each specific involved task and the correspondent parsed workflow DSL
source code.

It should be noted that this module is shared between
the FLOWViZ and the workflow system components.
\section{Discussion}
\label{discussion}
This paper introduces FLOWViZ - a tool integration framework for phylogenetic
frameworks. This provides both a client and a server component, allowing the
user to build workflows, through a user-friendly web client's interface,
enabling it to add its own phylogenetic tools and use them in its own built
workflows, by only requiring few lines of configuration.

This provides a great tool scalability and interoperability, as tools can be continuously integrated, while the workflow system's workers can be also scaled-out to handle larger workloads.
Tool interoperability is also supported by using contracts and loosely coupled relationship between components, which
allows seamless integration with other phylogenetic frameworks, requiring the
developer to only add the phylogenetic framework's bundled tools to the FLOWViZ
configuration.

The proposed architecture was tested and materialized into an application
prototype, composed by two main components: (i) a React web client and (ii) an
HTTP server, both written in JavaScript. With these two it is possible to: (i)
integrate external phylogenetic tools, (ii) build workflows with the previously
integrated tools and, finally, (iii) retrieve results and logs from the
workflow's execution.

\begin{figure}
    \centering
    \includegraphics[width=\linewidth]{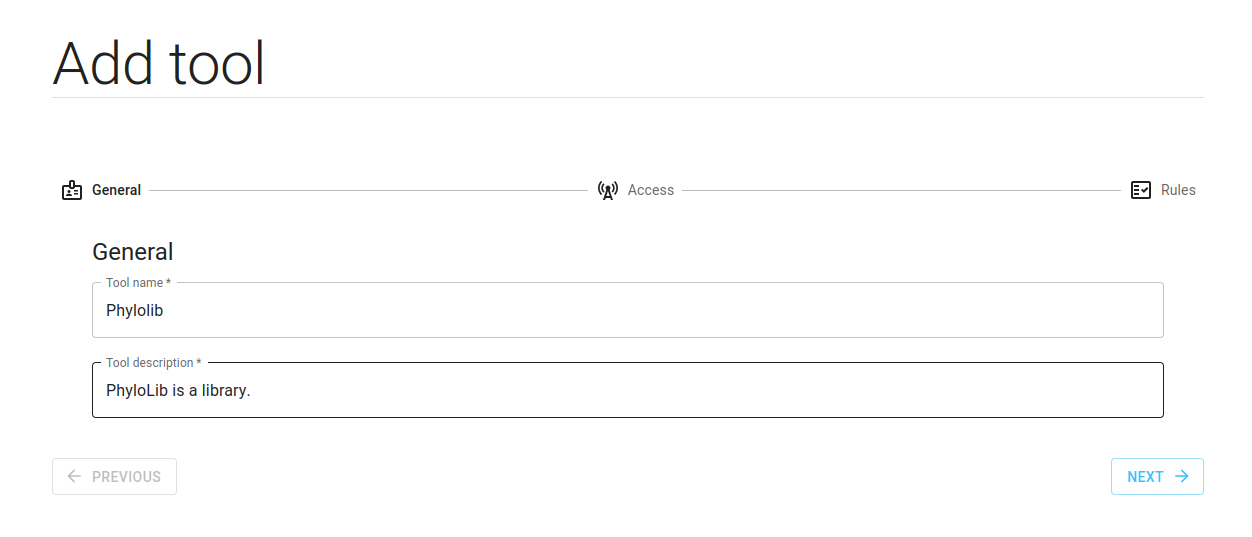}
    \caption{FLOWViZ: tool integration general fragment}
    \label{fig:toolGeneral}
\end{figure}

\begin{figure}
    \centering
    \includegraphics[width=\linewidth]{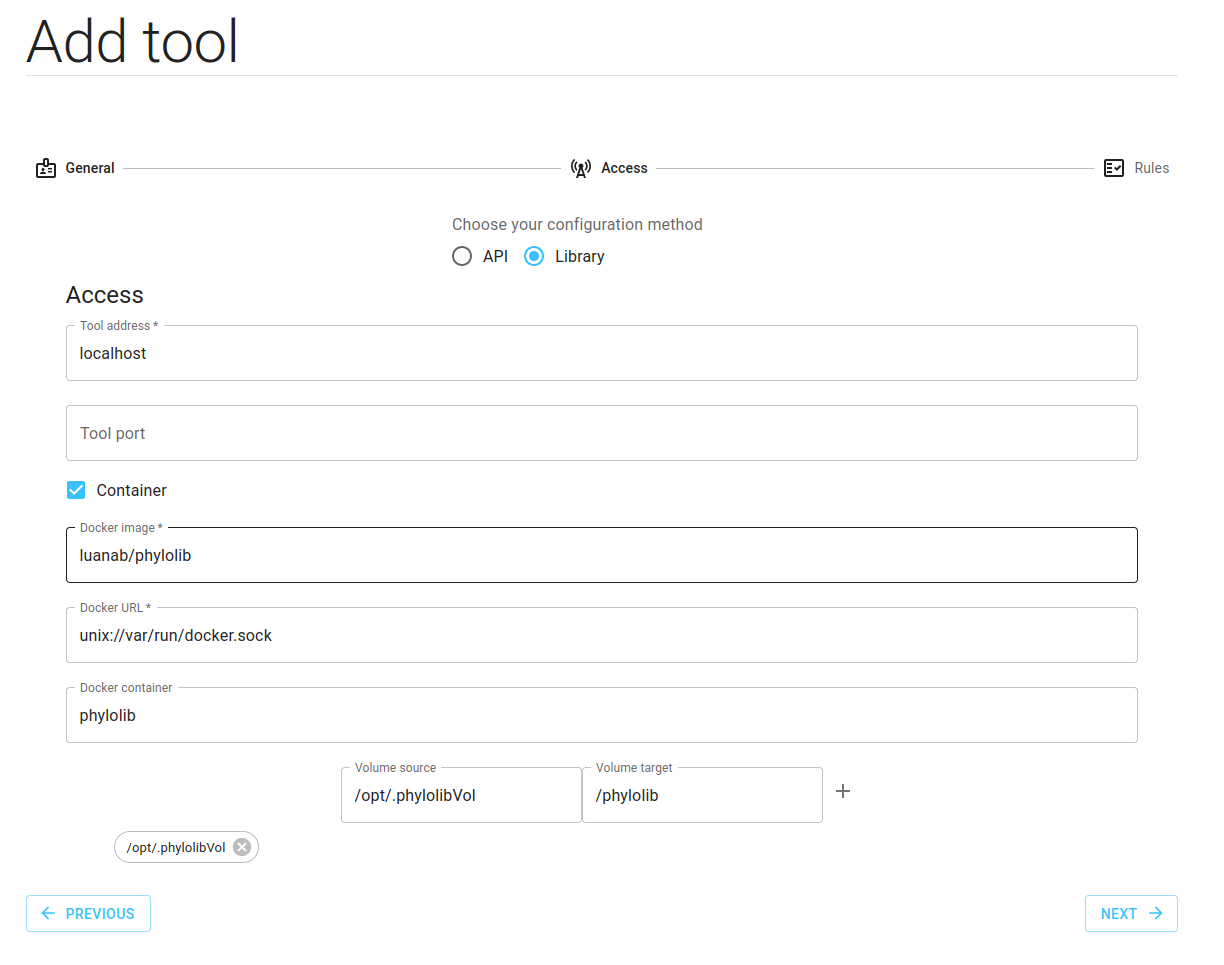}
    \caption{FLOWViZ: tool integration access fragment}
    \label{fig:toolAccess}
\end{figure}

\begin{figure}
    \centering
    \includegraphics[width=\linewidth]{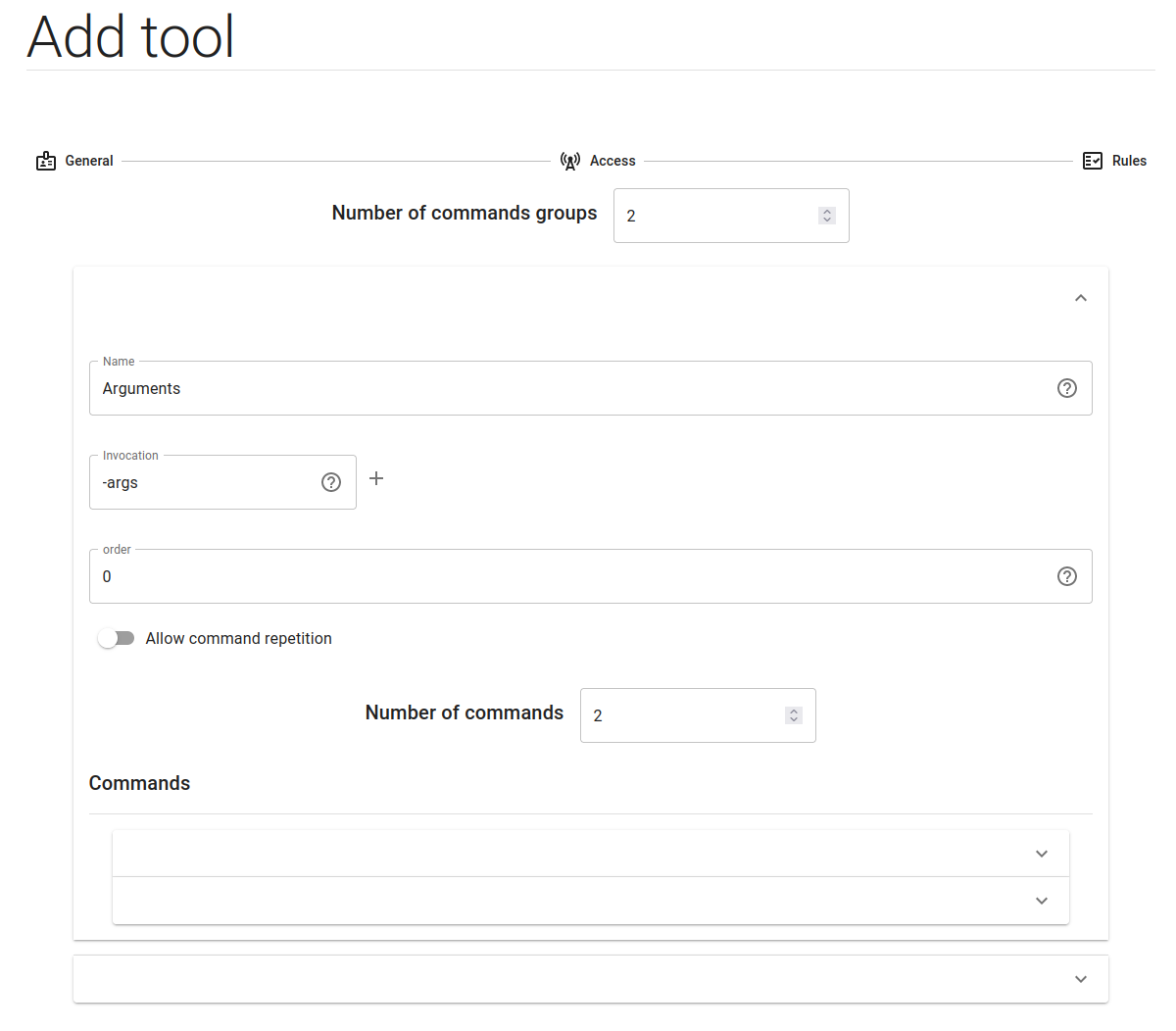}
    \caption{FLOWViZ: tool integration rules fragment}
    \label{fig:toolRules}
\end{figure}

Figures \ref{fig:toolGeneral}, \ref{fig:toolAccess} and \ref{fig:toolRules} show the tool integration use case implementation in the application.

\begin{figure}
    \centering
    \includegraphics[width=\linewidth]{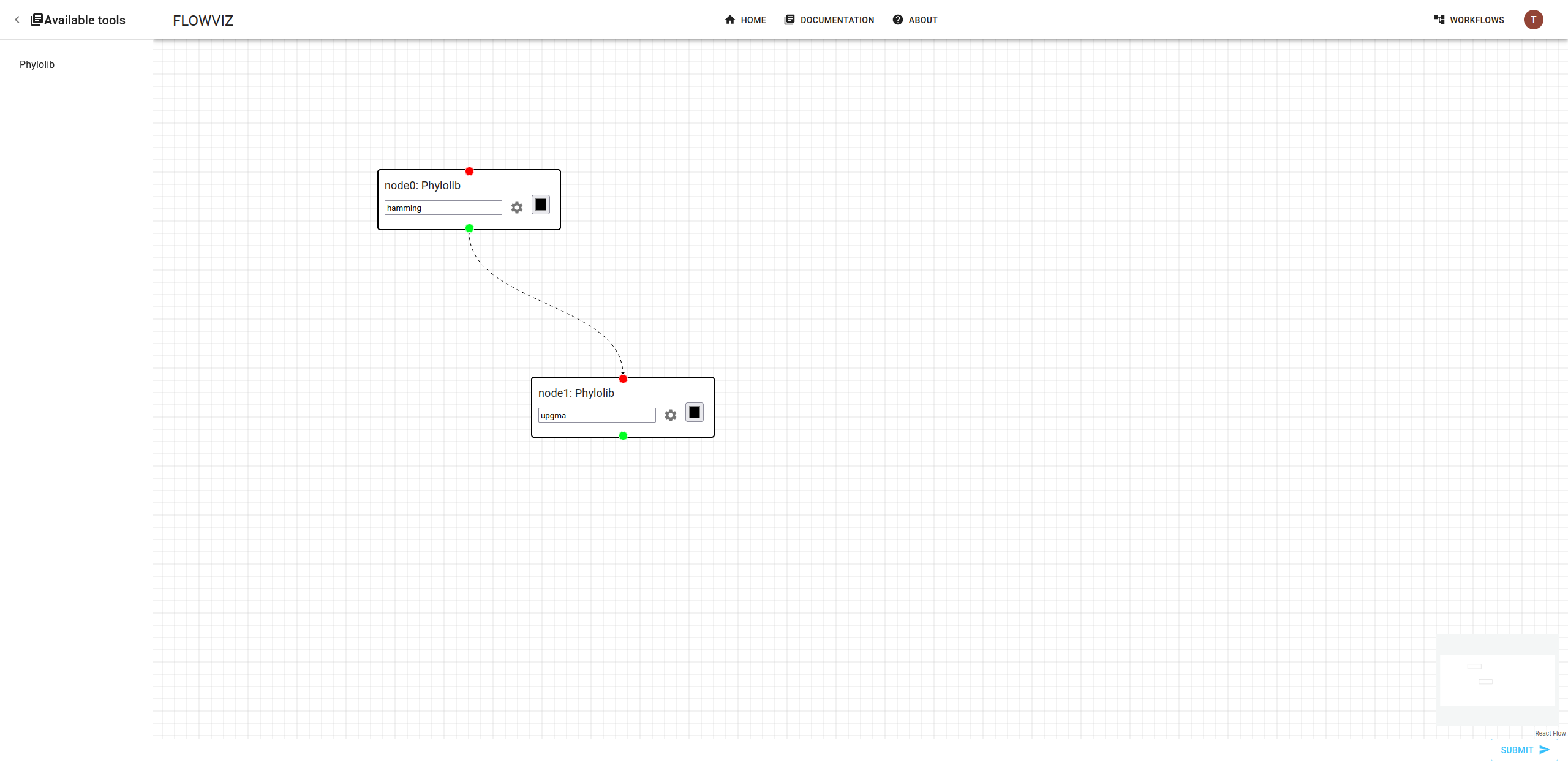}
    \caption{FLOWViZ: \textit{whiteboard}}
    \label{fig:whiteboard}
\end{figure}

After a successful tool integration, the web client will send the tool contract
to the HTTP server, which will go through a validation. If the contract is
valid, it will be saved into the database. At this point, the tool was
successfully integrated. The user can then use the previously integrated tools
to build its customized workflows, using the web client's provided editor - the
\textit{whiteboard}, where the user can graphically draw and configure the
workflow's tasks (figure \ref{fig:whiteboard}).

\begin{figure}
    \centering
    \includegraphics[width=\linewidth]{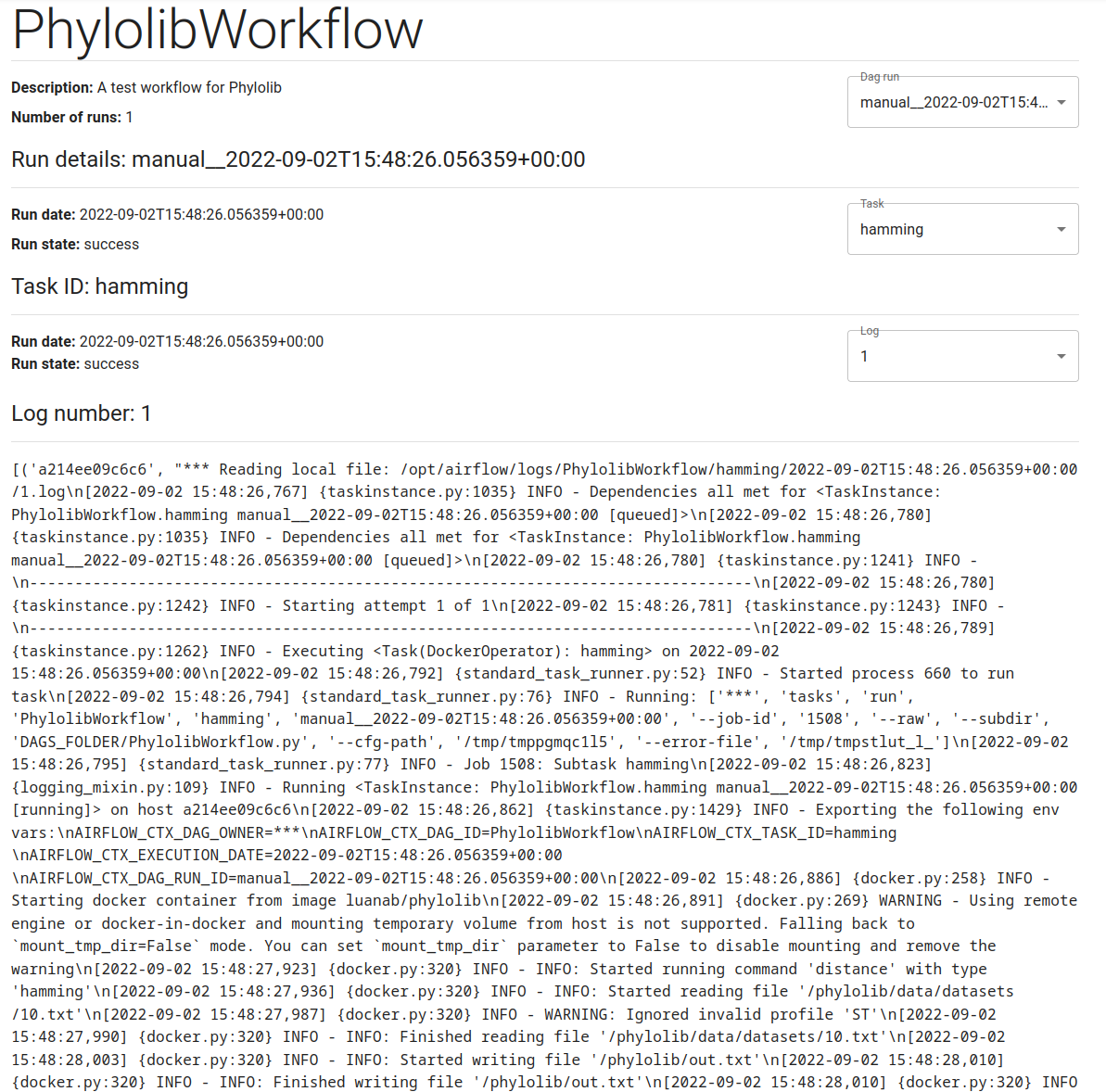}
    \caption{FLOWViZ: Results of a workflow execution}
    \label{fig:workflowResults}
\end{figure}

When the user finishes the configuration, the workflow can be submitted and send
to the HTTP server, which will go again, through a validation. If it succeeds,
the workflow data will be saved into the database and a HTTP request, containing
the name of the workflow and its correspondent user, will be sent to the
Airflow's REST API, that will then fetch from the database the correspondent
user workflow and parse it to an Airflow DSL script or DAG. The DAG will then be
executed at the pre-configured date and time, which will produce results, that
can be easily consulted via the web client (figure \ref{fig:workflowResults}).

\section{Acknowledgements}
The work reported in this article was supported by national funds through
Funda\c{c}\~ao para a Ci\^encia e a Tecnologia (FCT) with reference
UIDB/50021/2020 and project NGPHYLO PTDC/CCI-BIO/29676/2017. It was also
supported through Instituto Politécnico de Lisboa with project IPL/2021/DIVA.

\small
\bibliographystyle{plain}
\bibliography{bib}

\end{document}